\documentclass[a4paper,twocolumn,11pt,unpublished]{quantumarticle}
\pdfoutput=1
\usepackage[utf8]{inputenc}
\usepackage[english]{babel}
\usepackage[T1]{fontenc}
\usepackage{amsmath}
\usepackage{hyperref}
\usepackage[numbers]{natbib}

\usepackage{tikz}
\usepackage{lipsum}
\usepackage{amsmath}

\newcommand{\be}{\begin{equation}}
\newcommand{\ee}{\end{equation}}

\begin{document}

\title{PANSATZ: Pulse-based Ansatz for Variational Quantum Algorithms}

\author[1]{
 Dekel Meirom}
  \affil[1]{Faculty of Electrical Engineering,
  Technion - Israel Institute of Technology,
  Haifa 3200003, Israel}
  \orcid{0000-0002-8500-3838}
  \thanks{\texttt{dekelmeirom@gmail.com}}

 \author[2]{Steven H. Frankel}
  \affil[2]{Faculty of Mechanical Engineering,
  Technion - Israel Institute of Technology,
  Haifa 3200003, Israel}
  \thanks{\texttt{frankel@technion.ac.il}}

\maketitle
\begin{abstract}
 We develop and implement a novel pulse-based ansatz, which we call PANSATZ, for more efficient and accurate implementations of variational quantum algorithms (VQAs) on today's noisy intermediate-scale quantum (NISQ) computers.  Our approach is applied to quantum chemistry.  Specifically, finding the ground-state energy associated with the electron configuration problem, using the variational quantum eigensolver (VQE) algorithm for several different molecules. We manage to achieve chemical accuracy both in simulation for several molecules and on one of IBM's NISQ devices for the $H_2$ molecule in the STO-3G basis. Our results are compared to a gate-based ansatz and show significant latency reduction - up to $7\times$  shorter ansatz schedules. We also show that this ansatz has structured adaptivity to the entanglement level required by the problem.
\end{abstract}

\section{Introduction}

Today's quantum computers (QCs) are often described as noisy intermediate-scale quantum (NISQ) computers due to the relatively low numbers of qubits available (e.g. 10's to 100's) and the relatively high levels of noise associated with them (e.g. decoherence and gate fidelity errors) \cite{NISQ, NISQ_challenges_opportunities, NISQ_review}. These limitations result in circuits with short width and shallow depth.  

To make use of these NISQ machines, a class of hybrid quantum-classical algorithms are being developed that seek to leverage the relative strengths of quantum and classical computers.  The most common example of such algorithms are variational quantum algorithms (VQAs) \cite{VQA_review}.  VQAs use a QC to prepare a short-depth parameterized quantum circuit (PQC) representing a trial solution or ansatz to the problem at hand. Measurements of a final quantum state are used to calculate a cost function, which is then minimized on a classical computer to estimate the problem solution. Prominent examples of VQAs include the variational quantum eigensolver (VQE) for quantum chemistry and materials applications \cite{VQE_first_paper, VQE_HEA2017}, quantum approximate optimization algorithm (QAOA) for combinatorial optimization problems \cite{QAOA}, and variational quantum linear solver (VQLS) for linear algebra problems  \cite{VQLS}.

One of the main challenges associated with effective VQA implementations is related to the design of a suitable PQC/ansatz that balances expressibility and noise, while avoiding exponentially vanishing gradients of the cost function, referred to as the barren plateau (BP) problem. The two main categories of ansatz that have been considered include problem-inspired ansatz (PIA) and hardware-efficient ansatz (HEA) \cite{VQE_HEA2017}. PIA structure is primarily determined by the details of the problem being solved.  HEA structure is determined by the properties of the target hardware. HEAs are designed to reduce PQC depth while maintaining a general and expressive ansatz. There have been a number of attempts \cite{VQE_best_practices} to improve the gate-based ansatz approach including ADAPT-VQE \cite{ADAPT-VQE} and Noise-Adaptive Search (QuantumNAS) \cite{QuantumNAS}.

Recently, the idea of combining quantum optimal control (QOC), a method by which optimal pulses can be designed to improve qubit coherence and gate fidelity, with VQA has been proposed  \cite{pulses2circuits}.  A few recent studies have sought to implement this and related ideas, proposing to bypass the PQC at the gate level for a pulse-based state-preparation \cite{ctrl-VQE, ctrl-VQE_multi_levels}, ansatz generation \cite{PAN, QOC_ansatz}, machine learning tasks \cite{variational_pulse_learning} and better gate compilation using QOC techniques tailored for VQAs \cite{pulse_effecient_transpilation, pulse_efficient_QOC, CR_driven_VQA, DD_and_better_gate_calibration}.
In this paper, we propose an implementation of an ansatz at the pulse level, creating a new type of HEA that displays reduced latency and better expressibility with minimal additional parameters and optimization complexity. 

\subsection{Contributions}
We developed a general purpose parameterized pulse-based ansatz with a relatively small number of parameters. We demonstrated, using the VQE algorithm, that such an ansatz outperforms gate-based hardware efficient ansatz, which is widely used on today's NISQ devices. Using our ansatz, we were able to reach chemical accuracy on a real NISQ device. We believe that by using such a pulse-based ansatz, along with state of the art error mitigation techniques, this approach might bring us one step closer to solving real-world problems using quantum computers.

\section{Computational Approach}
\label{sec:comp}

\subsection{Gate-based ansatz}

Most of today's quantum computing logic is represented using quantum gates. The implementation of such quantum gates on quantum hardware is done using control pulses. In order to map between the desired logic of a quantum gate and a sequence of control pulses, a series of experiments that select the correct pulses has to be run on the hardware, which is usually referred to as the process of calibration. Today's QCs are noisy and unstable, which leads to the requirement of repeating calibrations frequently to maintain high-fidelity of the gate operations. 

\begin{figure}[htp]
    \centering
    \includegraphics[width=0.5\textwidth]{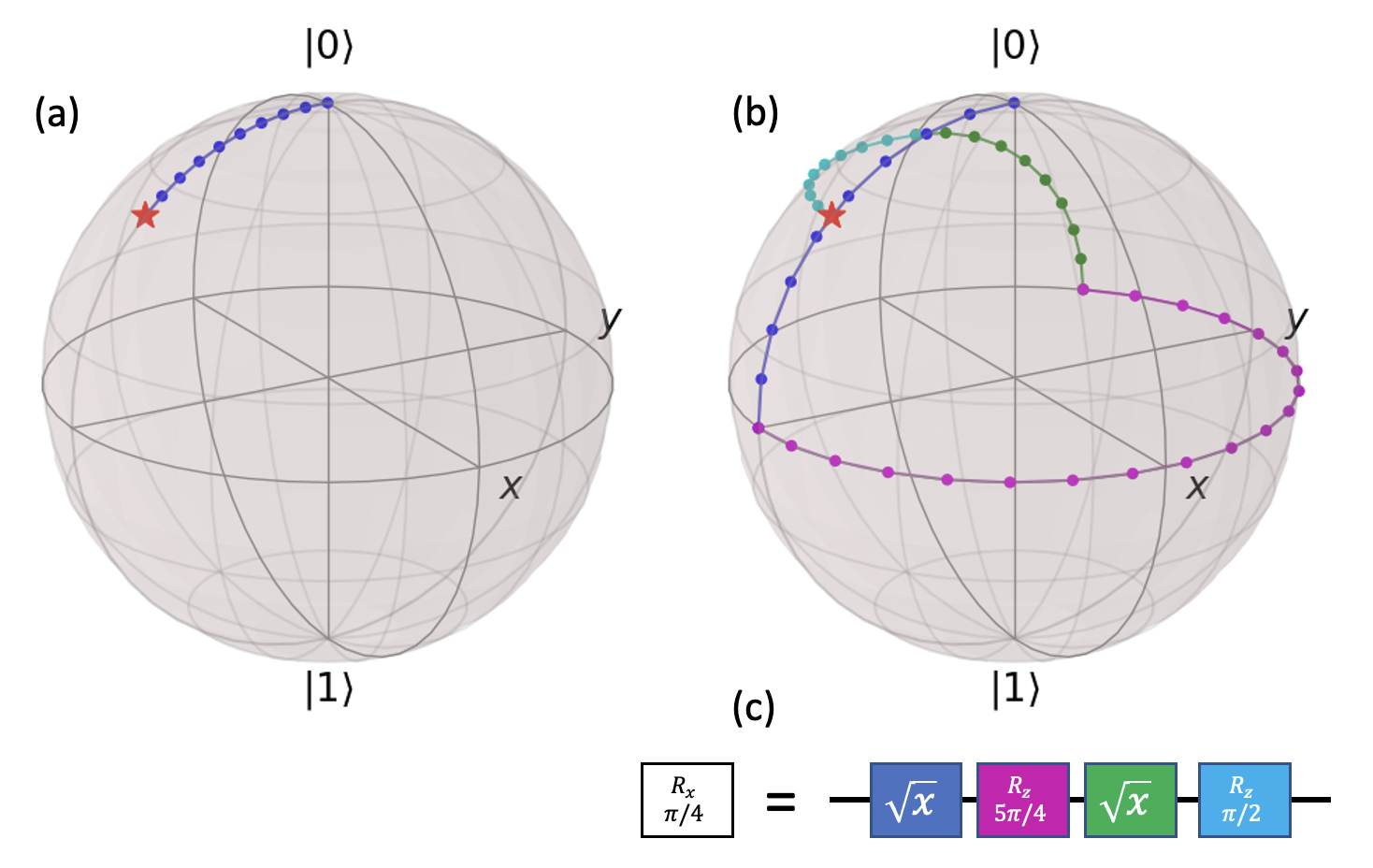}
    \caption{Demonstrating redundancy in the control pulses in non-native gate decomposition. (a) Qubit trajectory on the Bloch sphere which represents the logic of $R_x(\frac{\pi }{4})$ directly. (b) Qubit trajectory on the Bloch sphere of a decomposition of a $R_x(\frac{\pi }{4})$ gate into $R_z(\theta )$ and $\sqrt{x}$  gates. (c) The logical decomposition of $R_x(\frac{\pi }{4})$ gate into $R_z(\theta )$ and $\sqrt{x}$ gates.}
    \label{fig:pulse_vs_gate}
\end{figure}

The complexity of the calibration process and the need to do it very frequently limits the QC designers to only a small number of gates that will be mapped into control pulses. This set of gates should be universal (to allow universal computing) and is called the native gate set. Each quantum gate that is not included in the native gate set must be first decomposed into a sequence of native gates in a process termed transpilation. In most cases, such decompositions are not unique, and finding the optimal one is challenging.  The process usually introduces redundancy and added latency in the qubits manipulation compared to the case where the original gate was part of the native gate set and had its own mapping to control pulses. An example of this type of redundancy is shown in Fig.\ref{fig:pulse_vs_gate}.

The structure of the HEA tries to minimize the depth of the gates in their decomposed format in order to lower the noise of execution. Each HEA layer usually consists of an entangling layer, built from native two-qubit gates, followed by a parameterized single-qubit gate layer featuring rotation angles as parameters. In addition, there is another initial layer of parameterized single-qubit gates. Henceforth, we will refer to an ansatz built from gates a GANSATZ and use the Real Amplitudes Ansatz, which is commonly used in VQE problems, as a baseline for comparison to our new approach.

\begin{figure*}[htp]
    \centering
    \includegraphics[width=0.7\textwidth]{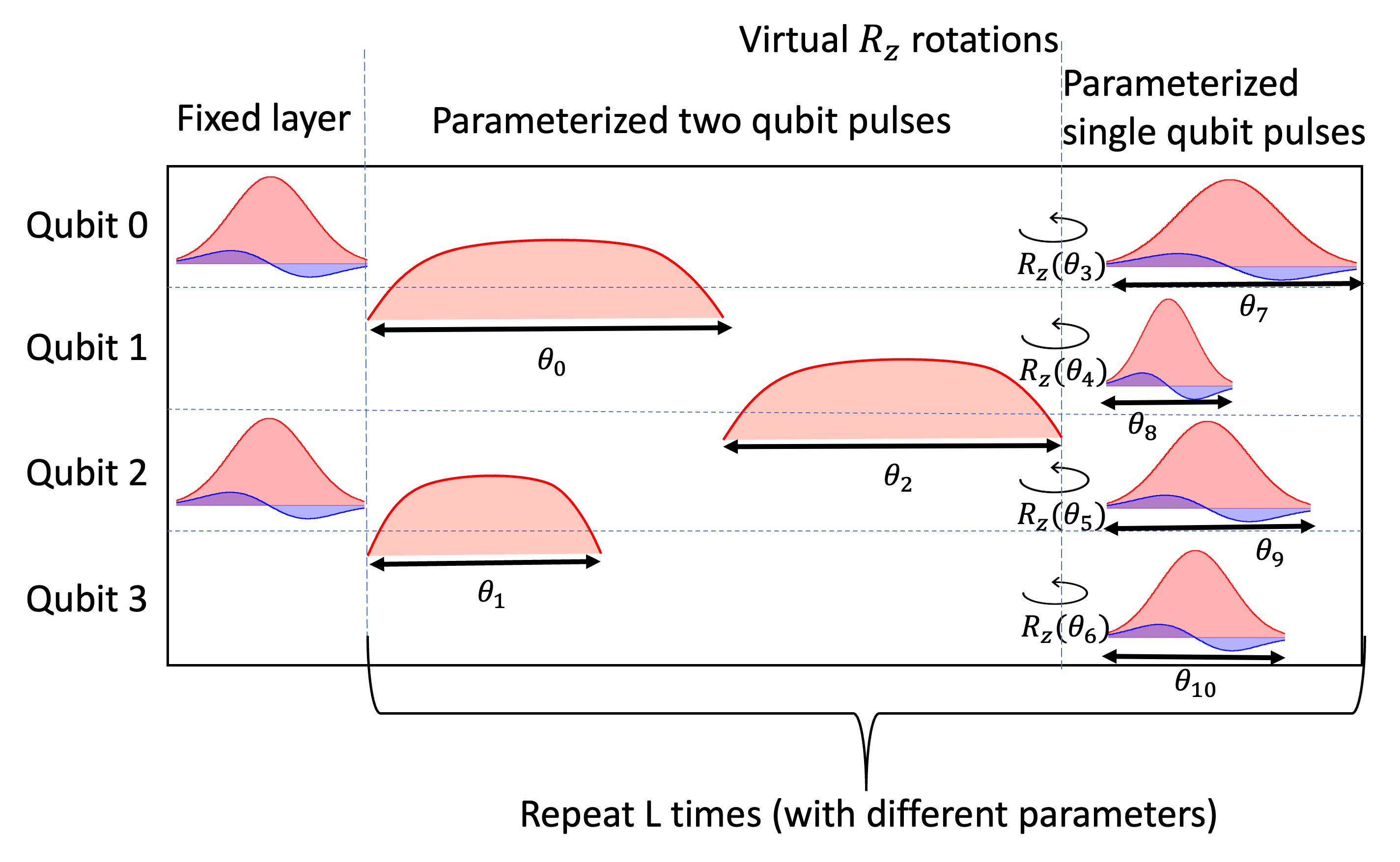}
    \caption{PANSATZ structure. The PANSATZ is built out from repeated $L$ layers, each consisting of parameterized two qubit pulses ordered in two alternating layers according to the device connectivity, followed by virtual $R_z$ gates and single qubit pulses on all qubits. The initial layer consists of fixed single qubit pulses which rotate the control qubits of the next layer to the XY plane. The two qubit pulses have flat-top Gaussian shape, and are coloured in light red, acting on the two qubits adjacent to the line it is drawn on. The single qubit pulses have DRAG shape, and are coloured in red (which represents the Gaussian part of the pulse) and blue (which represents the DRAG derivative part of the pulse).}
    \label{fig:pansatz_structure}
\end{figure*}

\subsection{Our approach: pulse-based ansatz}

The structure of our proposed pulse-based ansatz, which we will henceforth refer to as PANSATZ, is similar to the GANSATZ structure, where each gate is replaced with a parameterized pulse, including the two-qubit gates, which are also parameterized as part of the PANSATZ. As opposed to the GANSATZ, calibration is not performed and the pulses are not mapped into a logical unitary. Therefore, the logical unitary of the pulse sequences in the ideal case is unknown, and it is not trivial to trace the evolution of the qubits state after each pulse. Similar to other HEAs, the PANSATZ strives to prepare a quantum state while minimizing the incoherent noise associated with the preparation process (as VQAs are resilient to coherent errors \cite{VHQC_theory, VQE_error_robustness}, the focus is only on incoherent errors). By trading off knowledge about the unitary evolution in the noiseless case, the PANSATZ can have better expressibility (with the same number of layers), as each part of the unitary evolution can be parameterized, and lower incoherent noise compared to the GANSATZ.
Each pulse has many degrees of freedom which can be parameterized. While keeping more degrees of freedom as parameters improves the expressibility of the PANSATZ, it might also create over-parameterization and introduce trainability issues. Therefore, fixing some of the degrees of freedom as hyper-parameters and limiting the optimization process of the algorithm to only a few parameters for each pulse is crucial for the trainability of the ansatz. As the structure and properties of the control pulses vary between different quantum hardware, the choice of the parameters to optimize should be hardware dependent. Recently, a study focused on pulse level algorithms of natural atoms technology was published \cite{pulse_VQA_natural_atoms}.

\begin{figure*}
    \centering
    \includegraphics[width=0.95\textwidth]{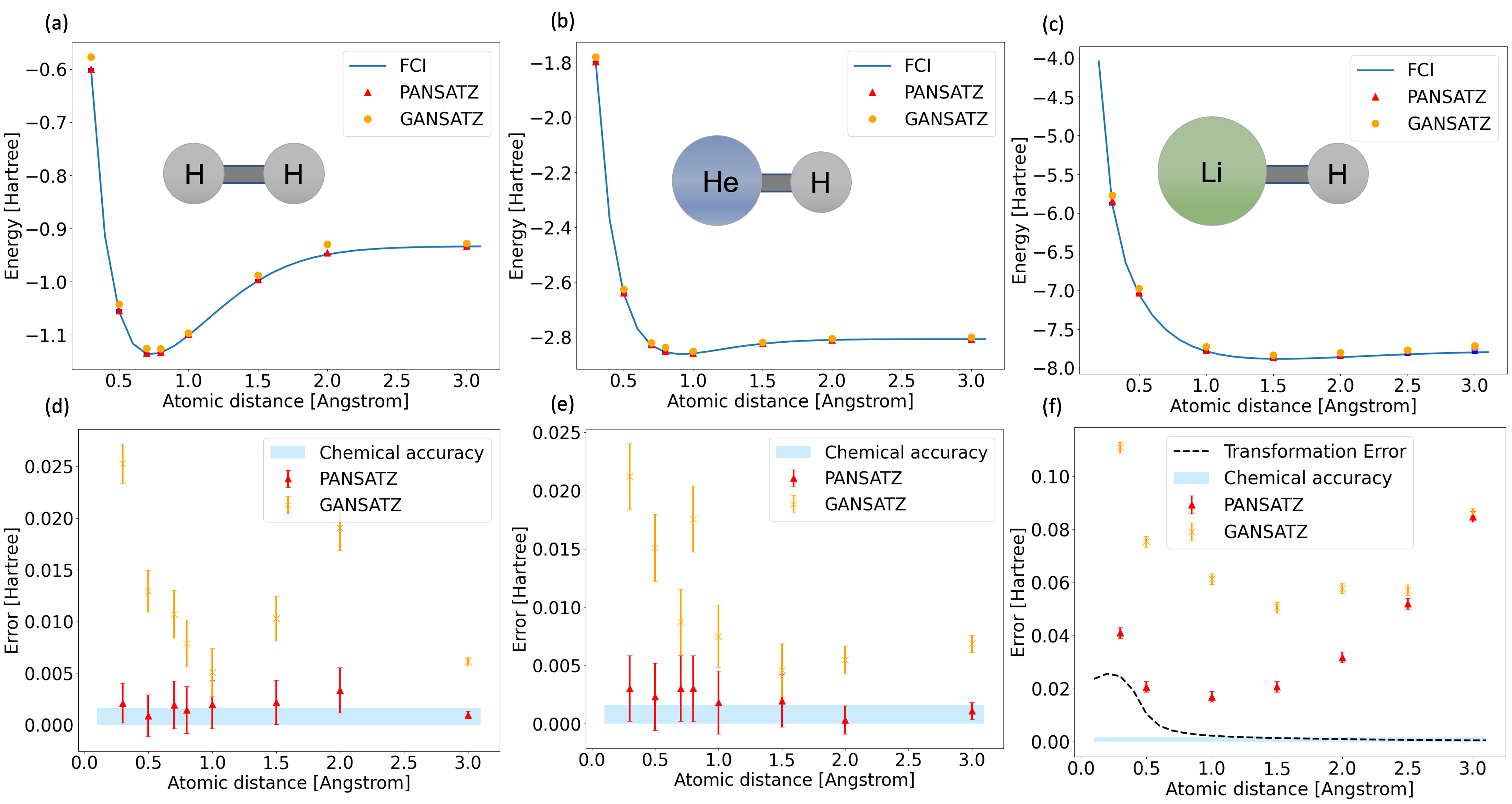}
    \caption{Simulations results. The top insets of each figure are representations of the molecular geometry, not drawn to scale. The error bars on the simulation data are smaller than the size of the markers. The reference on the FCI for each point is highlighted as blue tick marks. (a) Simulation results of the VQE on $H_2$ molecule. (b) Simulation results of the VQE on $HeH^+$ molecule. (c) Simulation results of the VQE on $LiH$ molecule. (d) Deviation of the simulation result from the FCI result of the VQE on $H_2$ molecule. (e) Deviation of the simulation result from the FCI result of the VQE on $HeH^+$ molecule. (f) Deviation of the simulation result from the FCI result of the VQE on $LiH$ molecule. The transformation error is the error introduced by reducing the number of qubits needed to encode the molecule's Hamiltonian and effectively reducing the active search space.}
    \label{fig:sim_res}
\end{figure*}

\subsection{Superconducting qubit implementation}

This study is focused on fixed-frequency superconducting qubits, primarily because of their availability on devices accessible via IBM cloud access. The driving Hamiltonian of each qubit for this hardware can be expressed in the rotating frame using the rotating wave approximation (RWA) as \cite{engineer_guide}:
\be
\begin{split}
H_c(t) = A(t) (e^{i((\omega_q - \omega_d(t))t + \phi(t))} \hat{a} + \\
e^{-i((\omega_q - \omega_d(t))t + \phi(t)} \hat{a}^{\dagger})
\end{split}
\ee
where $A(t), \omega_d(t), \phi(t)$ are the time-dependent amplitude, frequency, and phase of the driving microwave, $\omega_q$ is the frequency of the qubit and $\hat{a}^{\dagger}, \hat{a}$ are the bosonic creation and annihilation operators.
Previous studies have proposed to parameterize and optimize most of the possible driving Hamiltonian's time-dependent variables \cite{ctrl-VQE, PAN}. In our study, in order to avoid trainability issues and minimize the number of parameters, we propose to keep a fixed shape (the envelope of the waveform) to the pulses with only a few parameters to optimize, while the rest are calibrated as hyper-parameters at the start of the algorithm based on the given hardware.

Decoherence and dephasing are major sources of incoherent noise for this qubit architecture.  Hence, the duration of a quantum schedule dramatically affects the amount of noise in the computation. Therefore, \textit{pulse duration} was chosen as the optimization parameter (as illustrated in Fig.\ref{fig:pansatz_structure}).  This enables exploration of the Hilbert space with the shortest possible schedule duration. The pulse shape, driving frequency $\omega_d(t)$, amplitude and parameters associated with that shape, were chosen at the beginning of the algorithm as described below and kept fixed throughout the VQA computation. The phase $\phi(t)$ of the driving pulse is manipulated using $R_z$ gates, allowing the addition of only one parameter per qubit per ansatz layer, instead of an added parameter for each pulse envelope.

In our implementation, we choose derivative removal by adiabatic gate (DRAG) \cite{DRAG1, DRAG2} as the shape of the single-qubit pulses to minimize leakage into higher energy levels of the device. The $\sigma$ and $\delta$ parameters of the DRAG pulse were taken from the calibrated $X$ gate of the device and kept fixed. We also use flat-top Gaussian functions as the shape of the cross-resonance (CR) \cite{CR_pulse} pulses to enable smooth waveforms with maximum amplitude for most of the pulse duration. The $\sigma$ and the rise-fall ratio of the flat-top Gaussian are taken from the CR part of the CNOT gate of the device and kept fixed.
The frequencies of the pulses were chosen as the resonance frequency of the qubit they are controlling and the amplitude was chosen in a quick calibration process to find the maximum amplitude for short pulses which creates negligible leakage. Negative duration was allowed in order to account for unconstrained optimizers, while negative time was translated into negative amplitude and the absolute value of the duration. Between the layer of the CR pulses and the single qubit DRAG pulses, a layer of $R_z$ gates was added in order to enable changing the relative phase of each qubit, as the phase in the DRAG pulse is fixed. In superconducting qubits technology, such gates can be implemented as virtual gates with zero duration \cite{Virtual_Z_gate}, therefore they are perfect gates that do not inject any noise to the system.
We chose to keep the initial layer of single qubit pulses fixed with pulses that rotate some of the qubits to the XY plane of the Bloch sphere. The qubits that were chosen are those which will act as control qubits in the first entangling layer (which will be every other qubit if we assume linear nearest-neighbor connectivity).  This was done in order to create as much entanglement as possible from this layer, as the creation of entanglement is usually the most computationally time consuming.
Although the structure of the PANSATZ is fixed, by allowing some of the pulse duration parameters to be zero, some of the pulses and even entire layers can be kept out of the schedule (as opposed to the HEA, where the CNOT gates are not parameterized), adjusting the structure of the ansatz during the optimization process, benefiting from an adaptive approach similar to ADAPT-VQE \cite{ADAPT-VQE}.

\begin{figure}[htp]
    \centering
    \includegraphics[width=0.5\textwidth]{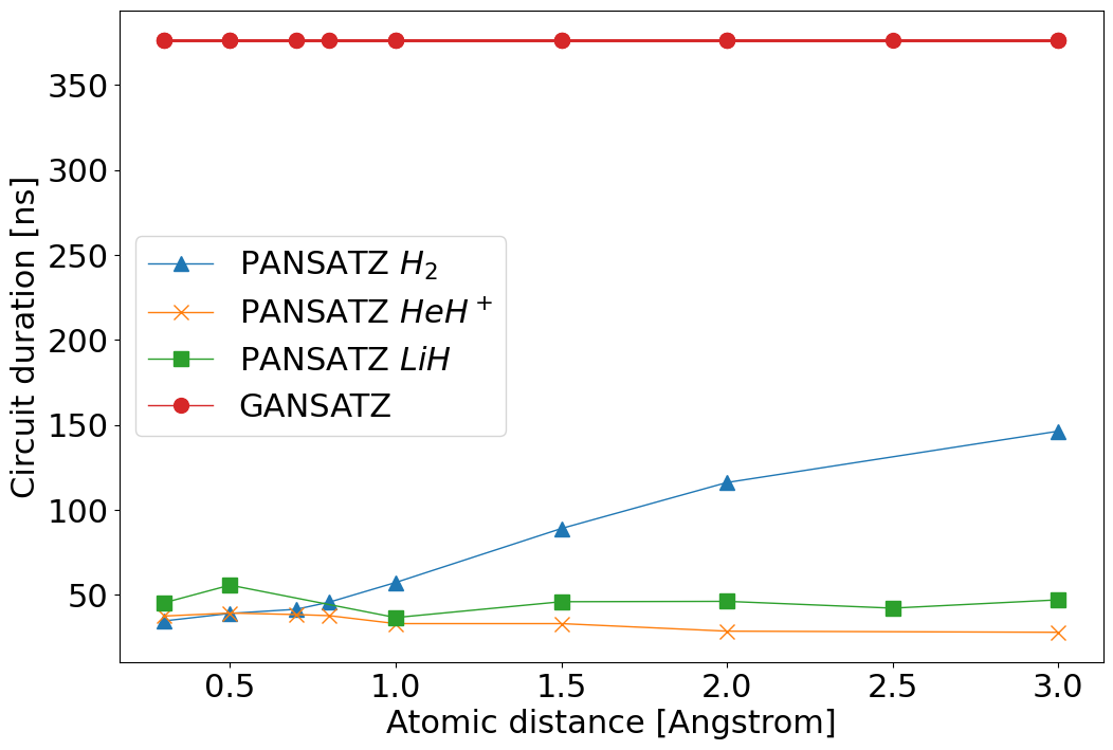}
    \caption{Duration of the PANSATZ schedules compared to Real Amplitudes HEA for $H_2, HeH^+$ and $LiH$ molecules, based on numerical simulations. The duration of the GANSATZ is constant because the ansatz has the same number of gates for each atomic distance, and the duration of each gate is fixed.}
    \label{fig:sim_dur}
\end{figure}

\subsection{Two qubit pulse implementation}
The two qubit pulses are usually the most important, as they can create entanglement between the qubits, but also the most time-consuming and error-prone. Therefore, additional care should be taken in the optimization process of these pulses.
The effective two-level system Hamiltonian describing the CR pulse can be approximated as the time-independent Hamiltonian \cite{open_pulse}
\be
\begin{split}
H_{CR} =  \omega_{IX} IX + \omega_{IY} IY + \omega_{IZ} IZ + \\
\omega_{ZI} ZI + \omega_{ZX} ZX + \omega_{ZY} ZY + \omega_{ZZ} ZZ
\end{split}
\ee
Where the dominant terms are the entangling term $ZX$ and the single-qubit terms $ZI$ and $IX$ \cite{CR_hamiltonian_coeff, CR_interactions, CR_active_cancellation}. The $ZX$ term is crucial for creating entanglement and is usually the only desired term of the CR pulse when designing a CNOT gate, while the $ZI$ and $IX$ are usually referred as unwanted terms and have the largest coefficients. In the PANSATZ case, there is no target unitary for the CR pulse, only a search for the mapping between the parameters of the pulse and the cost function. 
In our numerical experiments, we observed that a relatively small change in a CR pulse duration caused a large change in the cost function, mainly as a result of these heavy weighted single-qubit terms. These frequent large changes in the cost function harm the flow of the classical optimizer and lead to problems with the convergence of the algorithm. Therefore, we changed the CR pulses into an echo format \cite{CR_echo, CR_active_cancellation}, which has a DRAG-shaped flip pulse (which needs to be calibrated \textit{a priori}) on the control qubit between the two halves of a CR flat-top pulse. Such an echo cancels these single-qubit terms while keeping the entangling term. This makes each entangling pulse longer than it needs to be in the optimal case, but helps with the training process. This issue might be resolved by creating a tailor-made optimizer, but this is beyond the scope of this research.

\section{Results}
\label{sec:results}
In order to test our method we implemented the standard VQE algorithm but with our PANSATZ as the ansatz to find the ground state energy of different molecules. We first implemented the algorithm as a simulation, using the \texttt{Qiskit-dynamics} package \cite{qiskit_dynamics}, which utilizes the JAX array library to enable accelerated GPU execution, to find the minimum energy of the $H_2$, $HeH^+$ and $LiH$ molecules. Next, we 
implemented the algorithm on one of IBM's devices, \texttt{ibm\_lagos}, to find the minimum energy of $H_2$. All of the molecular Hamiltonians used in this work were computed in the STO-3G basis using \texttt{Qiskit-nature} package. These Hamiltonians were converted into spin Hamiltonians using parity transformation, and encoded into qubits after utilizing symmetries in the Hamiltonian, to reduce the amount of qubits needed in order to encode the solution to the problem. The $H_2$ and the $HeH^+$ Hamiltonians were represented using 2 qubits, while the $LiH$ Hamiltonian was represented using 4 qubits, which was achieved by reducing the active search space of the spin orbitals, as explained in \cite{VQE_HEA2017}. Such reduction introduces an error, especially in small atomic distances, but reduces the number of required qubits to encode the Hamiltonian significantly, which allows reaching better results.
We compared our results to the full configuration interaction (FCI) result with respect to the minimal STO-3G basis calculated by diagonalizing the Hamiltonian.
In both the simulation and on the actual IBM device hardware we used $10000$ shots for each circuit run to measure the expectation value with a small variance. In the run on \texttt{ibm\_lagos} we used tensored readout error mitigation \cite{readout_error_mitigation, readout_error_mitigation2}.
For these tasks, we used PANSATZ with only 1 layer, consisting of fixed single-qubit pulses, followed by parametrized two-qubit pulses and parametrized single-qubit pulses (as described in the previous section). This structure generates 5 parameters for the $H_2$ and $HeH^+$ molecules (while GANSATZ has 4 parameters) and 11 parameters for the $LiH$ molecule (while GANSATZ has 8 parameters). As initial pulse parameters, we chose duration for the single qubit pulses so the final state will be approximately the Hartree-Fock (HF) state. All the two-qubit pulses were initialized with zero duration (as there is no entanglement at the HF state). For example, the HF state for the $H_2$ molecule is the $|01\rangle$ state, so all the parameters were initialized with zero duration except the single qubit pulse of the first qubit. This pulse was initialized with the same duration as the pulse in the fixed layer (which is calibrated to create $~\frac{\pi}{2}$ rotation around the X-axis), resulting in a $\pi$ rotation to the first qubit, yielding the HF state.
We achieved good agreement between the simulation results and the results from \texttt{ibm\_lagos} device (after readout error mitigation). 

\begin{figure}[htp]
    \centering
    \includegraphics[width=0.5\textwidth]{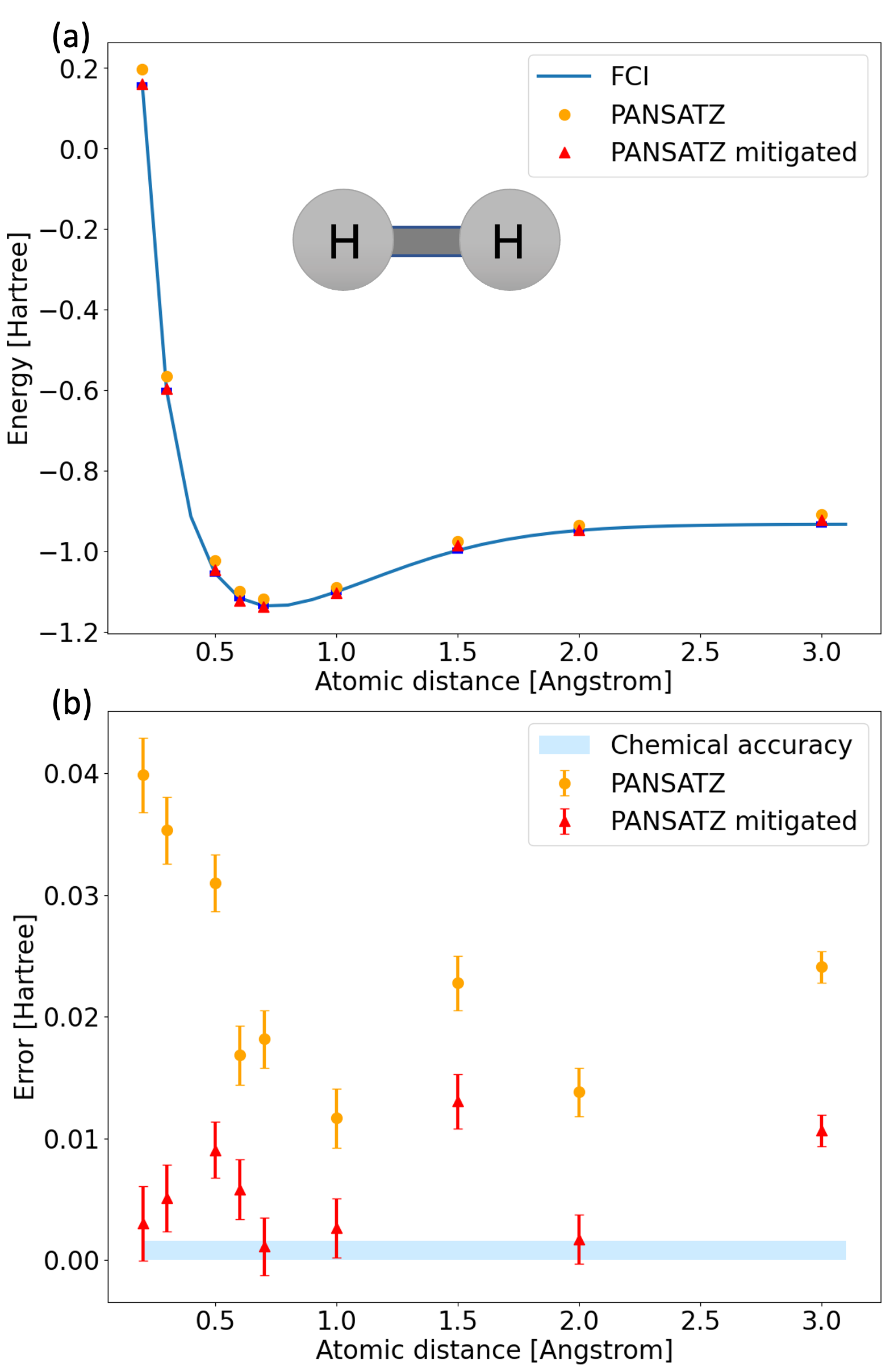}
    \caption{VQE results for $H_2$ molecule on $ibm\_lagos$ device. (a) The final ground energy found by the algorithm with and without readout error mitigation. The error bars on the data are smaller than the size of the markers. The reference on the FCI for each point is highlighted as blue tick marks. (b) Deviation of the result from the FCI result with and without readout error mitigation.}
    \label{fig:hardware_results}
\end{figure}

\subsection{Simulation}
We simulated fixed-frequency transmon qubits by using the following device Hamiltonian -
\be
H = \sum_{k=1}^{N} (\omega_k \hat{a}_k^{\dagger} \hat{a}_k - \frac{\delta_k}{2} \hat{a}_k^{\dagger} \hat{a}_k^{\dagger} \hat{a}_k \hat{a}_k) + \sum_{<kl>} g (\hat{a}_k^{\dagger} \hat{a}_l + \hat{a}_l^{\dagger} \hat{a}_k)
\ee
where $\omega$ is the qubit frequency, $\delta$ is the qubit anharmonicity, $g$ is the neighboring qubit coupling strength and $ \hat{a}_k^{\dagger}, \hat{a}_k$ are the bosonic creation and annihilation operators. For the simulation, we used the values reported by IBM on \texttt{ibm\_manila} for these parameters. In order to make the Hilbert space finite, we truncated the energy levels of the transmons at 3 levels to be able to take into account leakage to the $|2\rangle$ level. We also simulated relaxation and decoherence noise using the following dissipators:
\be
D_0 = \sqrt{\Gamma _0} \cdot \sigma ^+
\ee
\be
D_1 = \sqrt{\Gamma _1} \cdot (\hat{a}_k \hat{a}_k^{\dagger} - \hat{a}_k^{\dagger} \hat{a}_k)
\ee
where $\sigma ^+$ is the Pauli ladder operator ($\sigma _x + i\sigma _y$). We used $\Gamma _0, \Gamma _1$ which are proportional to $T_1 = T_2 = 100 [us]$. At the end of each calculation of qubit evolution, we sampled the final qubit state with $10000$ shots in order to introduce shot noise caused by finite sampling to the simulation.
For comparison we also ran a GANSATZ simulation with 1 layer using the IBM simulator. we used the same device parameters and connectivity and inserted relaxation and decoherence noises of $T_1 = T_2 = 100 [us]$, and simulated shot noise by sampling the results with $10000$ shots. In the gate model simulator leakage can not be simulated, so we assumed that there was no leakage.
The duration parameter can be seen either as a continuous parameter, or as a discrete parameter, with the smallest time unit of the classical controller of the quantum processing unit as the discrete unit (about $0.222 [ns]$ for most of today's IBM QCs). Therefore, we tested two different optimizers in the simulation - simultaneous perturbation stochastic approximation (SPSA), which is used for continuous parameters and approximates the gradient of the parameter vector using only 2 measurements regardless of size \cite{SPSA}, and steepest-ascent hill climbing \cite{hill_climbing, hill_climbing_search}, which is used for discrete parameters. Both optimizers had similar accuracy, while the steepest-ascent hill climbing converged with much fewer iterations. 
As shown in Fig. \ref{fig:sim_res}, using PANSATZ we reached chemical accuracy ($0.0016$ Hartree) compared to the FCI result, up to a standard deviation (calculated from the variance in the expectation value calculation), across all atomic distances studied for the $H_2$ and $HeH^+$ molecules. Because of the small number of parameters, this was achieved with only a few tens of iterations in the worst case. Throughout the optimization process of the parameters, the duration of the schedule slowly increased as entanglement was added to the prepared state by including also CR pulses (with non-zero duration) in the schedule. For cases where more entanglement was needed in order to reach chemical accuracy, the total duration of the schedule increased even further, until the optimizer converged to the correct result. For example, the entanglement of the ground state of the Hamiltonian of the $H_2$ molecule after the parity transformation is higher for large atomic distances. Therefore, as entanglement is usually the most time consuming to create in superconducting qubits, the optimizer converged at a longer schedule duration when solving these problems, as can be seen in Fig.\ref{fig:sim_dur}. Even the longest duration schedule achieved by using PANSATZ is less than half the duration of the equivalent GANSATZ.
For the $LiH$ molecule, 1 layer is not expressive enough for large atomic distances (although it is more expressive than 1 layer of GANSATZ). Therefore, the optimizer converged with low entanglement state, which encountered minimal noise due to the short duration and had better results than having longer CR pulses. Because of the low noise, good results were achieved in small atomic distances, but the algorithm solutions deviate from the FCI as the atomic distance increases.

Figure \ref{fig:sim_res} shows excellent agreement between the PANSATZ predictions of the ground state energy in Hartree units versus interatomic distance in angstroms to the FCI result.

\begin{figure}[htp]
    \centering
    \includegraphics[width=0.5\textwidth]{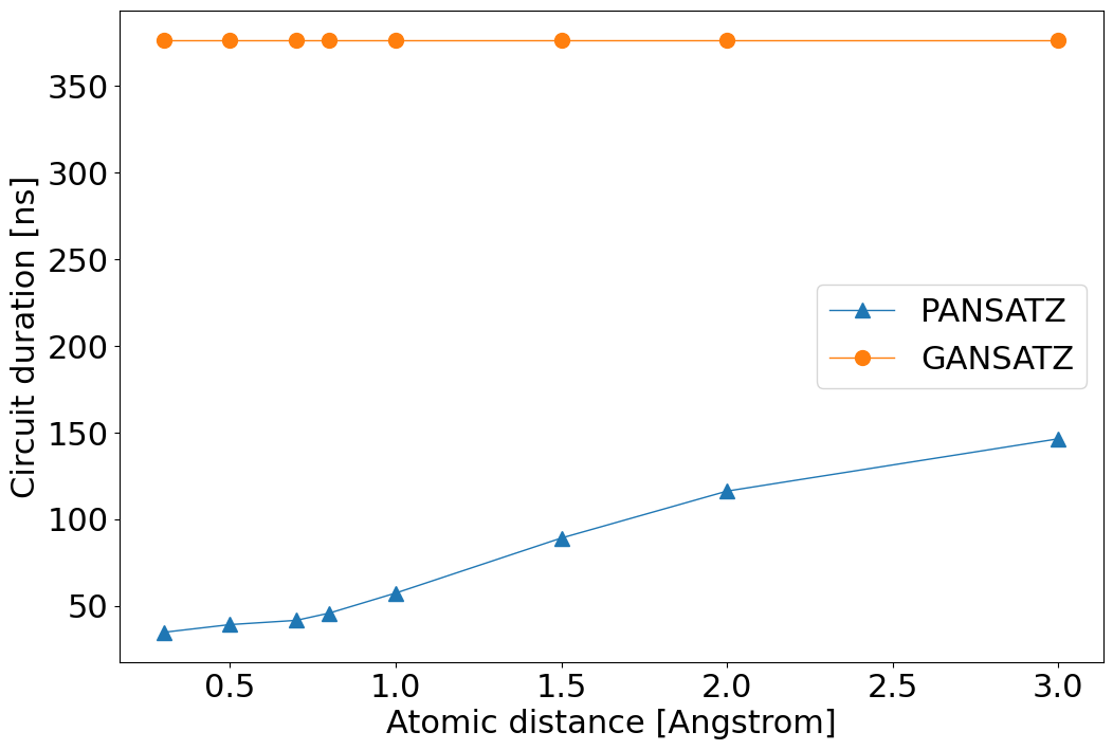}
    \caption{Duration of the PANSATZ schedules results from $ibm\_lagos$ compared to GANSATZ. The GANSATZ duration was taken as the duration of 1 layer of Real Amplitudes HEA transpiled on the same device}
    \label{fig:hardware_dur}
\end{figure}

\begin{figure}[htp]
    \centering
    \includegraphics[width=0.5\textwidth]{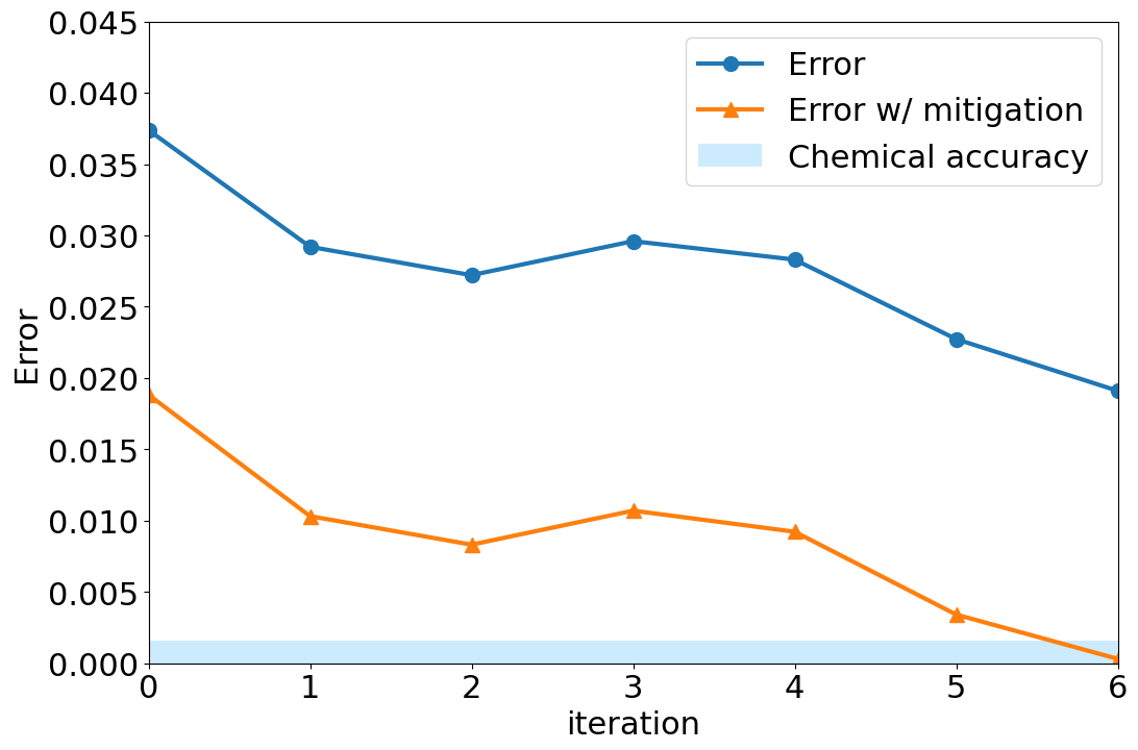}
    \caption{Convergence plot of the steepest-ascent hill climbing optimization algorithm for solving VQE for $H_2$ molecule with 0.7 angstrom atomic distance, run on $ibm\_lagos$.}
    \label{fig:hardware_convg}
\end{figure}

\subsection{Real Hardware}

We used \texttt{ibm\_lagos}, one of the IBM Quantum Falcon processors, to find the ground energy of the $H_2$ molecule at various atomic distances. We used the open-pulse feature \cite{open_pulse} to create the algorithm ansatz at the pulse level, and the PANSATZ structure described above. We used the steepest-ascent hill climbing algorithm for the classical optimizer, as it converged with fewer iterations in our simulations. We also used uncorrelated readout error mitigation in our post processing of the hardware measurement results, which is a scaleable method that can be used also when solving larger VQA problems. The results are shown at Fig.\ref{fig:hardware_results}. The results were obtained in a single convergence process, while convergence was declared by either reaching chemical accuracy or reaching 30 iterations. The readout error mitigated results show excellent agreement with the FCI results, including multiple points which reached chemical accuracy. These results were achieved as a direct result of the short duration of the PANSATZ schedule, as shown in Fig.\ref{fig:hardware_dur}, which was adaptive to the amount of entanglement needed for each atomic distance. The GANSATZ duration was taken as the duration of 1 layer of Real Amplitudes HEA compiled on the same device. By using a relatively small amount of parameters, a discrete optimization algorithm, and initial parameters which create a state close to the HF state, the PANSATZ reached the desired solution within only a few iterations (example shown in Fig.\ref{fig:hardware_convg}), giving hope for convergence within a reasonable amount of iterations also in larger problems.

\section{Summary and Discussion}

We developed a parameterized pulse-based ansatz, which we call PANSATZ, to be used with VQAs.  We chose pulse duration as the lone parameter. We tested PANSATZ in the context of VQE to find ground state energies of small molecules on both simulation and actual IBM hardware.  We achieved state of the art results, reaching chemical accuracy with the raw expectation value results (mitigating only readout errors), which, to the best of our knowledge, is an achievement that has not been shown yet on superconducting quantum hardware. Previous demonstrations of chemical accuracy on superconducting quantum hardware \cite{chemical_accuracy, purification} used extensive post-processing error mitigation techniques such as zero noise extrapolation (ZNE) \cite{ZNE_and_PEC, ZNE2}, probabilistic error cancellation (PEC) \cite{ZNE_and_PEC} and purification \cite{purification}, which might be used also with PANSATZ to make the results even more resilient to noise and help solve larger problems within the required accuracy. Adjustment of such mitigation techniques to make them suitable for pulses (for example, using pulse stretching instead of global folding in ZNE \cite{ZNE_analysis}) is left for further research.
Our experiments show significant latency reduction, resulting in improved accuracy of PANSATZ over typical gate-based ansatzes. The PANSATZ structure enables on-the-fly adaptation of the schedule latency depending on the entanglement level required to solve the given problem, potentially enabling simulations of larger molecules accurately. Our PANSATZ approach can be used with other VQAs as an improvement to the HEA. The use of PANSATZ in quantum algorithms where problem-inspired ansatze are used, such as QAOA, remains to be considered. Using a hybrid gate-based and pulse-based ansatz can be explored similarly to  \cite{HANSATZ_QAOA}.

\section{Acknowledgement}
The authors gratefully acknowledge the financial support of the Israel Science Foundation (ISF) on grant number 3457/21.
We would like to thank Dr. Adi Makmal for the fruitful technical discussions.
We also gratefully acknowledge the use of IBM Quantum services for this work and to advanced services provided by the IBM Quantum Researchers Program. The views expressed are those of the authors, and do not reflect the official policy or position of IBM or the IBM Quantum team.

\section{Data Availability}
The underlying code used in this study is available on GitHub and can be accessed via this link \url{https://github.com/dekelmeirom/PANSATZ}

\bibliographystyle{quantum}
\bibliography{pansatz} 

\end{document}